\documentclass[aps,prb,twocolumn,showpacs,reprint]{revtex4-1}

\usepackage{wrapfig}
\usepackage[]{graphicx}
\usepackage{tabularx}
\usepackage{booktabs}
\usepackage{textcomp}
\usepackage{amsmath}
\usepackage{amssymb}
\usepackage[]{graphics}
\usepackage{amssymb}
\usepackage{amsmath}
\usepackage{color}
\usepackage{ifpdf}
\usepackage{soul}

\newcommand{\executeiffilenewer}[3]{%
\ifnum\pdfstrcmp{\pdffilemoddate{#1}}%
{\pdffilemoddate{#2}} > 0 {\immediate\write18{#3}}\fi}

\ifpdf\usepackage{epstopdf}\fi

\begin{document}

\title{Optical properties of opaline photonic crystals covered by phase-change material Ge$_2$Sb$_2$Te$_5$}

\author{S.~A.~Dyakov}
\email[]{e-mail: s.dyakov@skoltech.ru}
\affiliation{Skolkovo Institute of Science and Technology, 143025 Moscow Region, Russia}

\author{M.~M.~Voronov}
\affiliation{Ioffe Physical-Technical Institute of the Russian Academy of Sciences, St. Petersburg 194021, Russia}

\author{S.~A.~Yakovlev}
\affiliation{Ioffe Physical-Technical Institute of the Russian Academy of Sciences, St. Petersburg 194021, Russia}

\author{A.~B.~Pevtsov}
\affiliation{Ioffe Physical-Technical Institute of the Russian Academy of Sciences, St. Petersburg 194021, Russia}

\author{I.~A.~Akimov}
\affiliation{{Experimentelle Physik 2, Technische Universit\"at Dortmund, 44221 Dortmund, Germany}}

\author{S.~G.~Tikhodeev}
\affiliation{A.~M.~Prokhorov General Physics Institute, RAS, Vavilova 38, Moscow, Russia}

\affiliation{Faculty of Physics, M.V. Lomonosov Moscow State University, 119991 Moscow, Russia}

\author{N.~A.~Gippius}
\affiliation{Skolkovo Institute of Science and Technology, 143025 Moscow Region, Russia}

\begin{abstract}
{The physical origin of resonant Wood's anomalies in the reflection spectra of {three-dimensional (3D)} opaline photonic crystals covered with Ge$_2$Sb$_2$Te$_5$ (GST225) is discussed. For this purpose, the optical reflection spectra are studied for different incident angles of light both experimentally and theoretically. The performed eigenmode analysis reveals that the Wood's anomalies originate from the quasiguided modes which appear in the GST225 capping layer. This conclusion is supported by the simulated electromagnetic near-field distributions of incident light at resonant frequencies. The experimental reflection spectra are in a good agreement with theoretical calculations performed by the Fourier modal method in the scattering matrix form.}
\end{abstract}


\maketitle 

\section{Introduction}

Opal-based materials attracted great attention of researchers over the several decades due to their unique properties, ease of fabrication and {high potential} for large variety of applications.  The optical properties of opals are strongly determined by the lattice configuration including its period, refractive index and orientation. Opals having photonic stop-bands in the visible range demonstrate shimmered interference colors. This fact not only provokes interest in opals from the viewpoint of jewellery but makes them promising for colorimetric sensors \cite{holtz1997polymerized, zhao2009encoded}, solar cells \cite{grandidier2011light, wehrspohn20123d, upping2011three} and microdisplays \cite{shim2010dynamic, lee2014controlled, yang2014metastable}. 

The {optical characteristics of the photonic crystals can be effectively tuned by mechanical deformation of the photonic crystal lattice \cite{yang2014metastable} or by applying the electrokinetic force for reshaping the opaline particles \cite{shim2010dynamic}. Another way to change the optical properties of opal is to control the geometry and composition of its surface. {Observation of surface states in pure 3D photonic crystals, without any specifically designed termination was described in literature both in dielectric \cite{ishizaki2009manipulation} and in plasmonic crystals \cite{tserkezis2011photonic}}. As it was demonstrated in Ref.\,\onlinecite{ding2010three}, the presence of gold film  terminating the opaline photonic crystal lattice leads to modification of angular transmission diagram owing to the coupling of the diffracted light to surface plasmon polaritons. In Refs.\,\onlinecite{dyakov2012surface, li2013surface, li2013tunable} it was shown that the unstructured silicon interfacial layer between the air and two-dimensional photonic crystal can give rise to Tamm-like surface states that modify the optical reflection of such structure in the spectral region of the photonic stop-band. Very recently, the optical Tamm states were demonstrated in transmission and reflectance spectra of three-dimensional opal photonic crystals coated by thin metal films \cite{korovin2016unconventional}. These Tamm states provide the bypass for light at the edges of the Bragg diffraction resonances and reduce the diffraction efficiency. Besides of that, the configuration of opal surface was demonstrated to play an important role in the second harmonic generation \cite{zaytsev2014enhancement} and the point imaging by opal slab with negative refraction \cite{xiao2004influence}.

The authors of Refs.\,\onlinecite{pevtsov2007ultrafast, voronov2014diffraction} described the optical properties of opal based structures impregnated or covered with materials which undergo a phase transition. The reflection spectra of such structures were considered in terms of resonant Wood's anomalies and 3D Bragg diffraction. {In Ref.\,\onlinecite{voronov2014diffraction}, the phenomenological theory was developed based on the Laue diffraction condition and the effective medium approximation. That theory satisfactory explained the angular and spectral positions of the resonant Wood's anomalies but was unable to describe amplitude of the peaks as well as the entire reflection spectra. Moreover, the physical origin of the discussed Wood's anomalies in opals remained unclear.} 
\begin{figure*}[t!]
\centering
\includegraphics[width=0.9\textwidth]{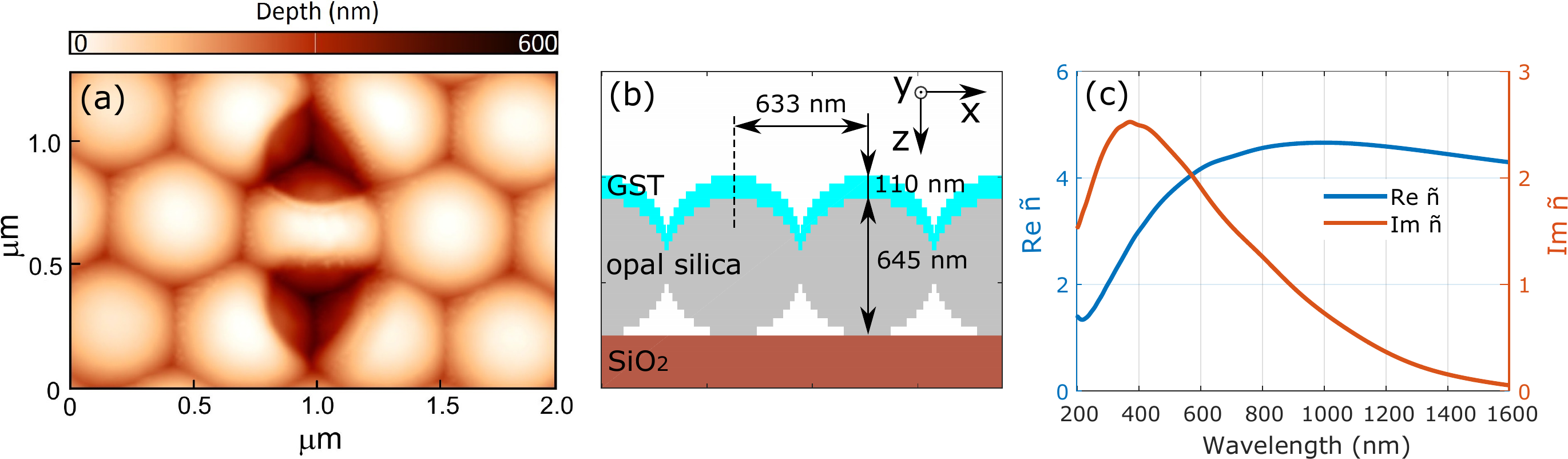}
\caption{(Color online) (a) AFM image of a fragment of the GST225/opal structure {where the GST225 film (light brown) is partially absent and SiO$_2$ spheres (dark brown) are exposed.} (b) Schematic view of the stucture used in scattering matrix method calculations. Only one monolayer of SiO$_2$ spheres is shown for simplicity. (c) Real and imaginary parts of GST225 refractive determined by ellipsometric measurements.}
\label{sample}
\end{figure*}

{The goal of the present work is to reveal the origin of the Wood's anomalies in the optical reflection of opals covered with high refractive index material Ge$_2$Sb$_2$Te$_5$ (GST225) \cite{raoux2014phase}. For this purpose we perform the eigenmode analysis, simulate the reflection spectra and calculate the electromagnetic near-field distribution in the opaline structures.}

The structure of the paper is as follows. In Sec.~\ref{sec:methods}, we describe our hybrid GST225/opal structures, their fabrication technique and the experimental setup. In Sec.~\ref{sec:calc}, we give a brief description of the scattering matrix method and the way how we approximate the structure to simulate our experimental results. In Sec.~\ref{sec:sim}, we present the results of reflection spectra measurements as well as the results of numerical simulations. We {analyse the angle dependencies of the Wood's anomalies and show how they relate to the dispersion relation of the quasiguided modes.} We also calculate the electromagnetic near-field distribution of {incident light. Finally,} in Sec.~\ref{sec:discussion} we propose a simple model in order to describe the  observed reflection spectra in terms of Lorentzian functions.

\section{Methods}
\label{sec:methods}
\subsection{Structure and experimental details}
The samples of the hybrid structures under study consist of 1 and 16 monolayers of amorphous silica spheres of diameter $\approx$645\,nm grown on fused silica substrate and covered by a GST225 film of $\approx$110\,nm thickness (see Fig.\,\ref{sample}a,b).  

{We determined the thicknesses of GST225 films deposited on opal by comparison with chalcogenide reference films deposited on a quartz substrate under the same technological conditions. The thicknesses of the reference films were measured by profilometer. These values are in reasonable agreement with the AFM data presented in Fig. 1a.}

The \textit{fcc} opaline lattice of spheres was grown from water suspension of {amorphous} SiO$_2$ particles by the vertical deposition technique \cite{jiang1999single}. A spread in diameter of spheres was smaller than 5\%. The amorphous GST225 layer was deposited on the surface of opal films in vacuum by the thermal deposition technique at a substrate temperature of 50$^\circ$C, an evaporator temperature of 600$^\circ$C, and a residual pressure of 10$^{3}$\,Pa (see Ref.\,\onlinecite{yakovlev2012controlling}). 

In optical measurements, the light of an incandescent lamp hits the structure at an incident angle $\theta$ changing from 11$^\circ$ to 66$^\circ$ with a step of 5$^\circ$. The azimuthal orientation of the samples was set using the 2D optical diffraction pattern from highly ordered (hexagonal) surface of the opal film. The following condition was satisfied: the incidence plane passed through one of the three equivalent pairs of sites of the hexagonal reciprocal lattice: (-11) (1-1); (01) (0-1); (-10) (10). The specular reflectance spectra were measured in the 900--1700 nm wavelength range with an Ocean Optics NIR 512 spectrometer equipped with an InGaAs-based charge-coupled device (CCD) array detector. Dependencies of refractive index, $\operatorname{Re}\left[\tilde{n}(\lambda)\right]$, and extinction coefficient, $\operatorname{Im}\left[\tilde{n}(\lambda)\right]$, of the GST225 films were determined by the spectral ellipsometer ``J.A. Woollam Co., Inc. model M-2000''.
\subsection{Method of calculation}
\label{sec:calc}
{Currently many powerful theoretical methods are available to describe the optical properties of periodically modulated layers \cite{mishchenko2002scattering, de2002retarded, moharam1995formulation, whittaker99, Tikhodeev2002b, ohtaka1996photonic, stefanou1998heterostructures}. In this work} the reflection spectra of hybrid structures are calculated using the {rigorous coupled wave analysis (RCWA) in the scattering matrix form  \cite{moharam1995formulation, whittaker99, Tikhodeev2002b}}. This method is based on splitting a structure into elementary planar layers, homogeneous in Z direction and 2D periodic in X and Y directions. In this splitting, the circular cross section of each sphere is approximated by a staircase. The solutions of Maxwell's equations for each layer are found by expansion of the electric and magnetic fields into Floquet-Fourier modes (plane waves). The exact solution can be presented as an infinite series over these modes, in the limit of infinite number of steps per each sphere. In numerical simulations, the scattering matrices are determined by taking a finite number of stairs per sphere diameter, $N_s$, and by truncating the Fourier series on a finite number of plane waves $N_p$. The calculation accuracy increases with an increasing $N_s$ and $N_p$, however, calculation time is proportional to $N_s$ and $N_p^3$ respectively. As a result, we used 13 plane waves in both X and Y directions so that the total number of plane waves was $N_g=169$. The number of stairs per diameter was $N_s = 25$ in what follows. Our test calculations revealed that at these $N_g$ and $N_s$ the computation scheme numerically converges.
\begin{figure*}[t!]
\centering
\includegraphics[width=\textwidth]{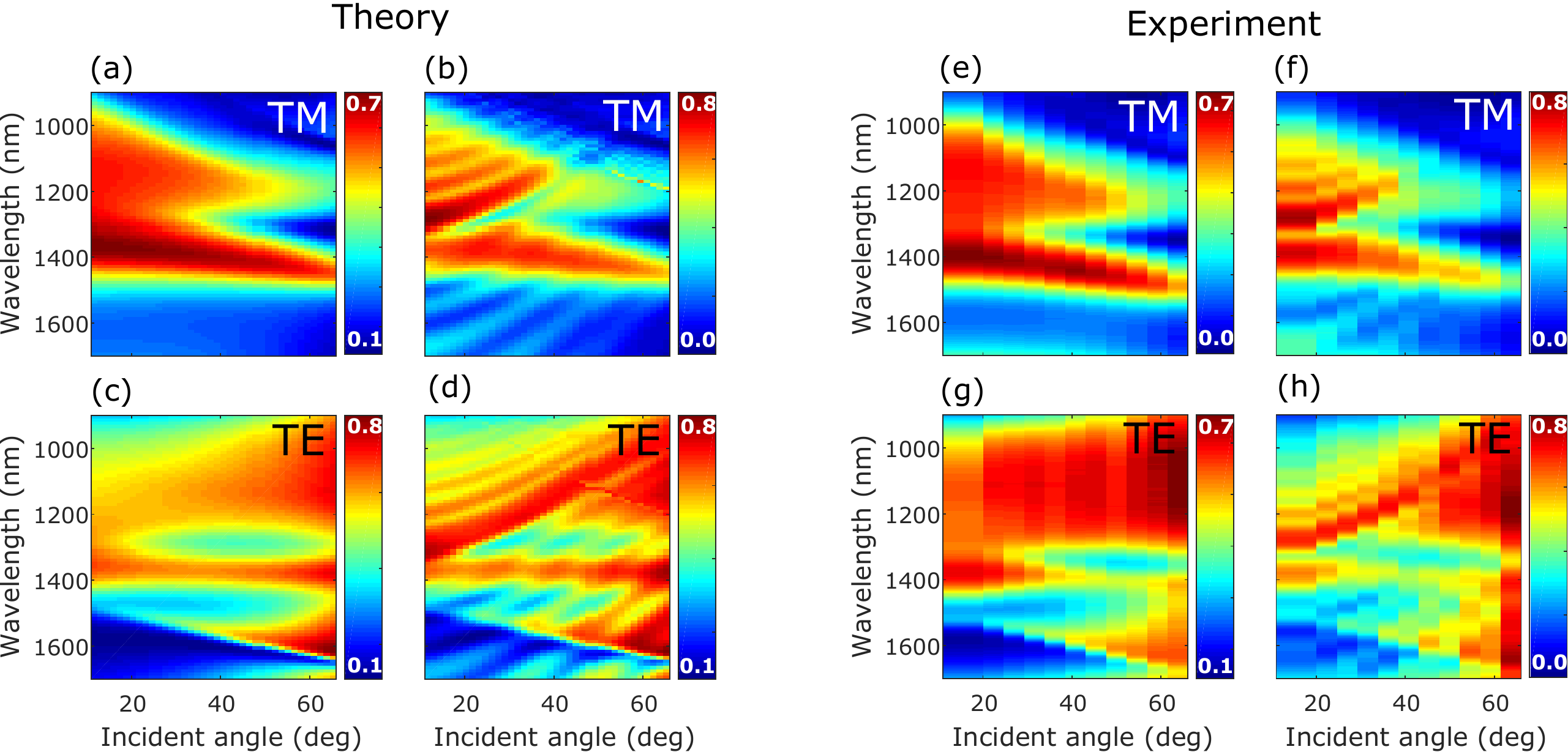}
\caption{(Color online) Theoretical (left panel) and experimental (right panel) reflection of GST225/opal structure versus wavelength and incident angle in TM and TE polarizations. The reflection coefficients are shown for one monolayer of silica spheres (panels (a),(c),(e) and (g)) and for 16 monolayers of silica spheres (panels (b), (d), (f) and (h)). The color scales are shown to the right of graphs. {The thickness of the GST225 capping layer is $h=110$\,nm for all panels.}}
\label{spectra}
\end{figure*}

\begin{figure*}[t!]
\centering
\includegraphics[width=0.8\textwidth]{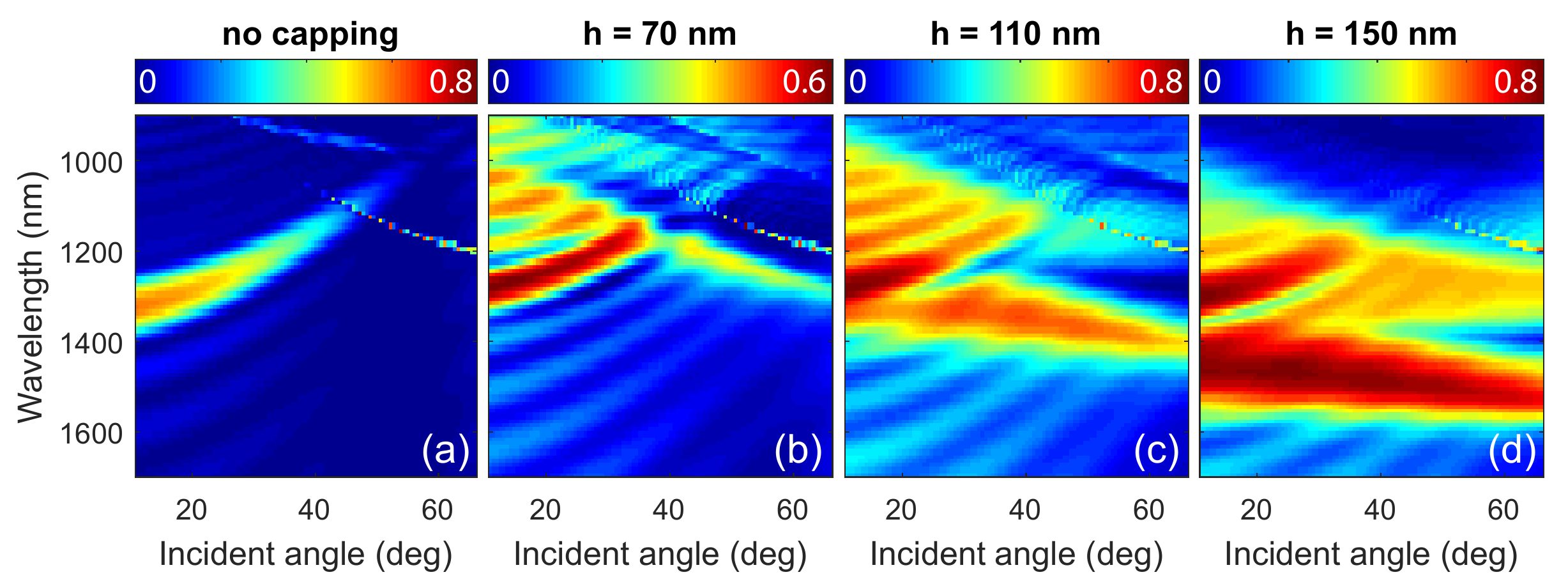}
\caption{(Color online) Comparison of calculated reflection spectra of GST225/opal structure for various GST225 film thicknesses: {h = 0, 70, 110, and 150 nm in panels a, b, c, and d, respectively} The color scales are shown on the top.}
\label{refh}
\end{figure*}
In addition, the choice of proper model of the geometry of GST225 capping layer is an important issue. It will be seen later that a so-called crescent model adequately describes the influence of the capping layer on the reflection properties of structures in study. 

The dispersion of refractive index of GST225 is measured using the ellipsometry technique and takes the imaginary part into account (see Fig.\,\ref{sample}c). Refraction index of opal spheres  is determined from  the analysis of Bragg reflection spectra of the initial opal film (denoted as opal silica in Fig.\,\ref{sample}b) and is taken as 1.92 and is assumed to be independent on the wavelength \cite{gajiev2005bragg}.

\begin{figure*}[t!]
\centering
\includegraphics[width=\textwidth]{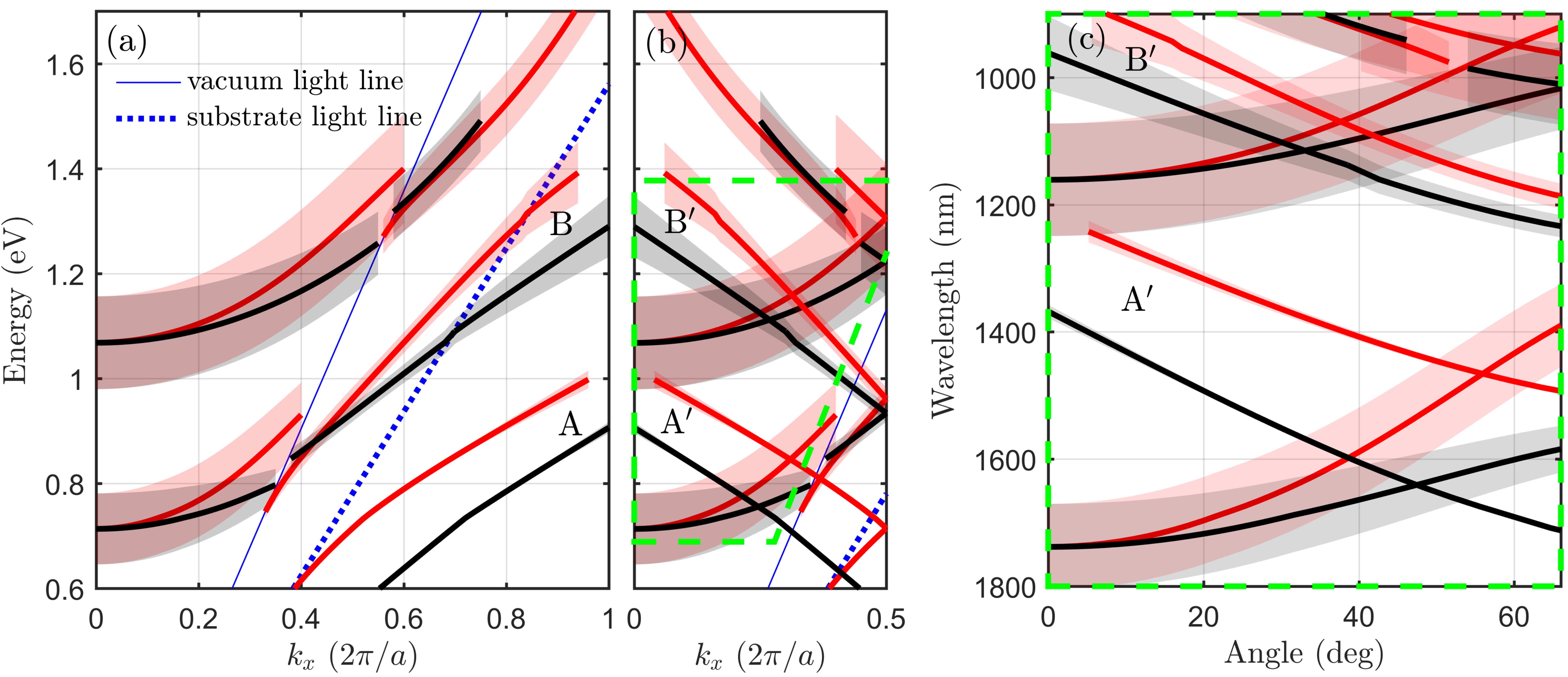}
\caption{(Color online) (a) The calculated dispersion relations of eigenmodes of hybrid GST/opal structure in empty lattice approximation. GST225 film thickness is 110\,nm. The red (black) color means the TM (TE) polarization. The width of red and black stripes in eV equals to $1/3$ of imaginary part of the corresponding {eigenenergy}. The blue solid and dotted line represent the vacuum and substrate light lines. {The vertical dashed line denotes the boundary of the first Brillouin zone.} (b) The dispersion folded in the first Brillouin zone. The dashed green line bounds the area which is represented in (c) in $\lambda-\theta$ coordinates. }
\label{disp}
\end{figure*}
\begin{figure}[t!]
\centering
\includegraphics[width=0.9\columnwidth]{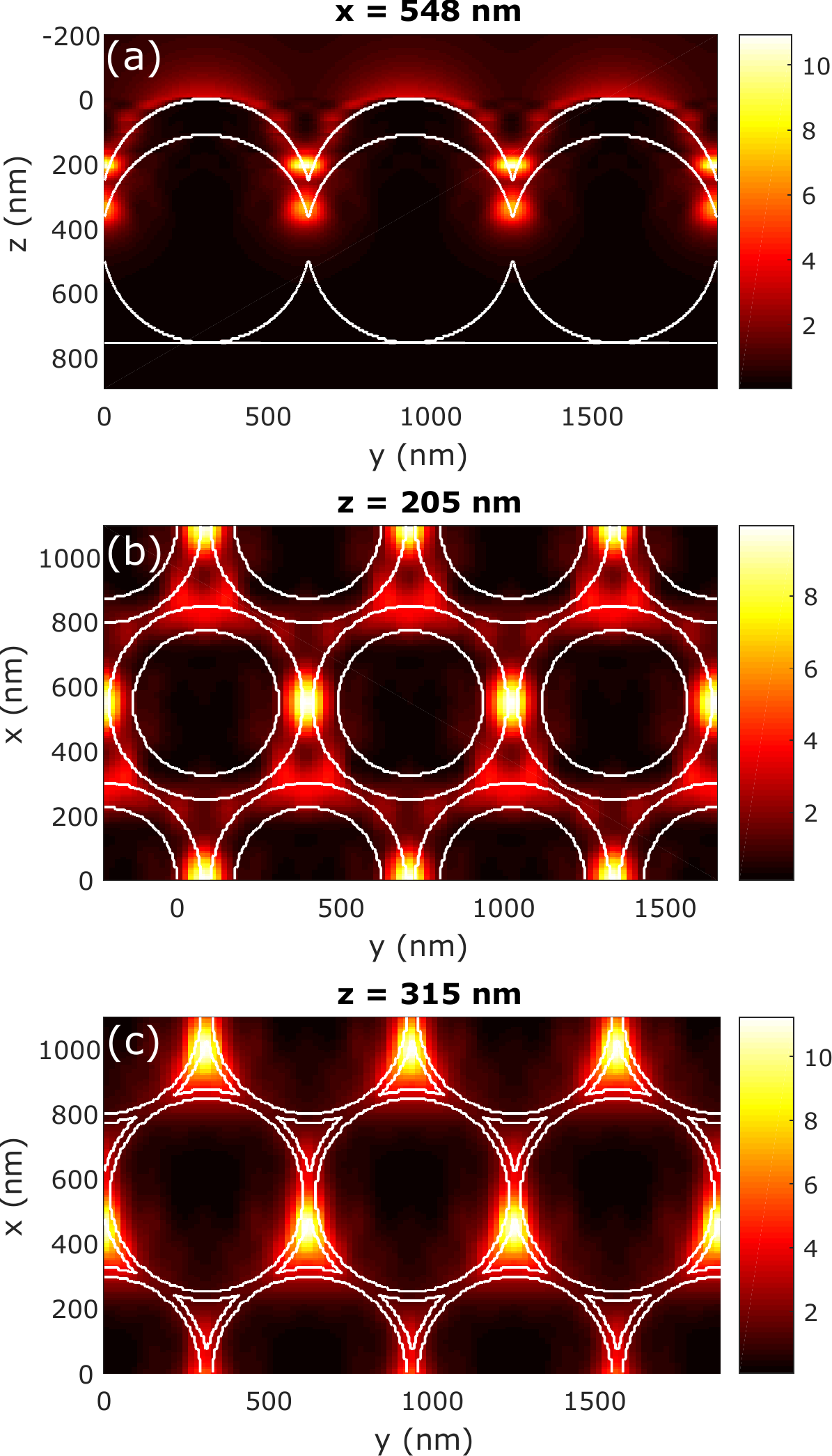}
\caption{(Color online)  Intensity of electric field of incident electromagnetic wave at $\lambda = 1310$\,nm, $\theta=56^\circ$ in TM polarization calculated in displayed cross-sections. White lines denote the interfaces between materials within the hybrid structure. {Dashed horizontal lines in panel a show the positions of cross-sections in panels b and c. Electric field intensity is normalized to that of incident plane wave}. The color scales are shown on the right.}
\label{fields}
\end{figure}

\section{Results}
\label{sec:sim}
\subsection{Reflection spectra}
The theoretical and experimental reflection coefficients of structures under study collected at different angles of incidence in TE and TM polarizations are shown in Fig.\,\ref{spectra} as two-dimensional function of photon energy and incident angle. In particular, Fig.\,\ref{spectra}a shows the theoretical reflection spectra of one monolayer of silica spheres covered by GST225 film. One can see that the spectra have two peaks which are redshifted with increase of the incident angle. {In Ref.\,\onlinecite{voronov2014diffraction} these peaks were attributed qualitatively to the Wood's anomalies corresponding to two different quasiguided modes. }  At small enough angles $\theta$ the reflection spectra are strongly deformed and take an asymmetrical form, so that it becomes impossible to differ between separate peaks in the interaction region. The value of the reflection coefficient at peak positions (for the light reflected in a small spatial angle) also monotonically increase with decreasing the angle $\theta$, reaching its maximum at normal incidence ($\theta = 0$). At very small angles $\theta$ the Wood's anomalies merge into a broad irregular contour. The similar behaviour is observed in TE polarization (see Figs.\,\ref{spectra}c,d). 

The  theoretical reflection spectra of 16 monolayer hybrid structure show up the peaks of the similar shape and the same spectral position (see Fig.\,\ref{spectra}b) along with the reflection maxima associated with photonic stop-band and the Fabry-P\'{e}rot interference over the structure thickness.  It is noteworthy  that the theoretical and experimental spectra (see Figs.\,\ref{spectra}e,g,f,h) are in a good agreement with each other. 

{In order to emphasize the influence of the GST225 film thickness, we calculated the reflection spectra of {the structure without capping and with different capping thicknesses} (see Fig.\,\ref{refh}). The reflection of the structure without capping layer does not show the Wood's anomaly described above (see Fig.\,\ref{refh}a). With increase of the GST film thickness, the Wood's peaks are redshifted while the photonic stop-band position and Fabry-P\'{e}rot resonances are practically not affected by GST225 film thickness. In addition to the photonic stop-band and Wood's anomalies, the spectra in Fig.\,\ref{refh}a-d show the sharp cusps which correspond to opening of the first diffraction channel (i.e. correspond to Fano-Rayleigh anomalies).}

\subsection{Dispersion of eigenmodes}
The analysis of the reflection spectra made in the previous section suggests that the reflection peaks such as those in Fig.\,\ref{spectra}a {are attributed to} the interface layer of the hybrid structure. In order to {reveal} the origin of electromagnetic resonances in greater detail, the poles of the scattering matrix as a function of complex frequency $\omega$ and {in-plane wavevector} $\left(k_x,k_y\right)$ have to be found \cite{Gippius2005c,Gippius2010,Weiss2011,Bykov2013}. The poles  define the energy dispersion of the eigenmodes $\omega(k_x,k_y)$. Here $k_{x,y}$ are the $x,y$ projections of the momentum vector of the incident wave. However,  it appears that in this particular case   the eigenmode dispersion can be found with a good accuracy from a simplified in-plane homogenized problem in the empty lattice approximation \cite{Tikhodeev2002b}. In this approximation, the dispersion relation is calculated for the new effective structure consisting of homogeneous isotropic layers corresponding to the consecutive steps of the staircase representation of the hybrid structure. The effective dielectric permittivities of these layers are calculated by the effective medium theory: 
\begin{equation}
\varepsilon_{eff} = \frac{\sum_\alpha\varepsilon_\alpha f_\alpha}{\sum_\alpha f_\alpha},
\end{equation}
where $\varepsilon_\alpha$ is the dielectric permittivity of $\alpha$-th component in a layer, $f_\alpha$ is the filling factor of $\alpha$-th component.

The approximate matrix $S$ can be represented in the form of a 2$\times$2 matrix. To account for the surface periodicity, the obtained dispersion curves are then folded back on the $\omega-k_x$ diagram into the first Brillouin zone.

The dispersion relations of eigenmodes of one monolayer of silica spheres covered by 110\,nm GST225 film are shown in Fig.\,\ref{disp}. Figure \ref{disp}a shows the dispersion curves of the effective non-periodic structure. The modes above the vacuum light line are the Fabry-P\'{e}rot resonances. They have relatively large bandwidth (imaginary part of eigenfrequency) {because of radiative losses} due to the energy leakage to the far field. {As to the modes below the substrate light cone (denoted as A and B), in the homogenized model they do not have radiative losses  at all. Their non-zero imaginary part of eigenfrequency is completely due to absorption losses in GST225. (In this respect, the homogenized model is valid provided that the radiative losses of modes A and B are smaller than the absorptive ones.)}  The non-zero imaginary part of eigenfrequency is originated from the light absorption in GST225. The modes A have longer lifetimes in comparison to the lifetimes of the modes B because the extinction coefficient of GST225, Im$[\tilde{n}]$, at 0.9\,eV is smaller than that at 1.35\,eV (see Fig.\,\ref{sample}c). In Fig.\,\ref{disp}, the stripes indicating the imaginary parts of eigenenergies of modes A are very narrow and are not visible to the naked eye.}

Let us now account for the periodicity and fold the dispersion curves shown in Fig.\,\ref{disp}a into the first Brillouin zone (see Fig.\,\ref{disp}b). The obtained eigenmodes A$'$ and B$'$ have now a negative dispersion. In order to compare the displayed dispersion curves with the experimental reflection spectra we plot the area bounded by the dashed green line in Fig.\,\ref{disp}b in $\lambda-\theta$ coordinates (see Fig.\,\ref{disp}c).

The comparison between the dispersion curves in Fig\,\ref{disp}c and the reflection spectra in Fig.\,\ref{spectra}a reveals that the TM reflection maxima at $\approx 1.4$\,eV and $\approx 1.2$\,eV are associated with the TM modes A and B. The other modes are not seen in the spectra as they have very large bandwidth. The correspondence between the reflection spectrum in TE polarization and the TE modes A and B is not so good. The reason for this is maybe a stronger optical strength of the resonances, stronger interaction between the quasiguided and Fabry-P\'{e}rot modes resulting in larger deviations from the empty lattice approximation \cite{Gippius2010}. 

{The above analysis of the structure eigenmodes enables us to conclude that the reflection peaks shown in Fig.\,\ref{spectra}a are associated with the resonant Wood's anomalies which correspond to quasiguided modes.}
  
\subsection{Field distributions}
In order to visualize the modes described above, let us  calculate the near-field distribution in the hybrid structures. We limit our discussion to the mode A, GST225 layer thickness $h = 110$\,nm and TM polarization. The spatial distribution of the electric field {intensity} in the {single-monolayer} structure covered by GST225 film is shown in Fig.\,\ref{fields} for $\lambda=1310$\,nm and $\theta=56^{\circ}$. These wavelength and angle of incidence correspond to the position of the dip in the reflection spectrum shown in Fig.\,\ref{spectra}a. It can be seen in Fig.\,\ref{fields}a that the electric field is localized in the near-surface region and represents mostly a standing wave. In the antinodes, {the field intensity reaches a maximum which is about 10 times} higher than the field of incoming wave. A field modulation coefficient, i.e. the ratio between the maximal and minimal electric field energies over the period of electromagnetic oscillation exceeds unity by two orders of magnitude. This is also characteristic of the standing wave of the surface mode.

The spatial distributions of electric field calculated in two horizontal cross sections (marked by dashed horizontal lines in Fig. 5a) are shown in Figs.\,\ref{fields}b,c. It can be seen, that the electric field is mainly localized in the  gaps between the silica spheres.

\section{Discussion}
\label{sec:discussion}
\begin{figure}[t!]
\centering
\includegraphics[width=1\columnwidth]{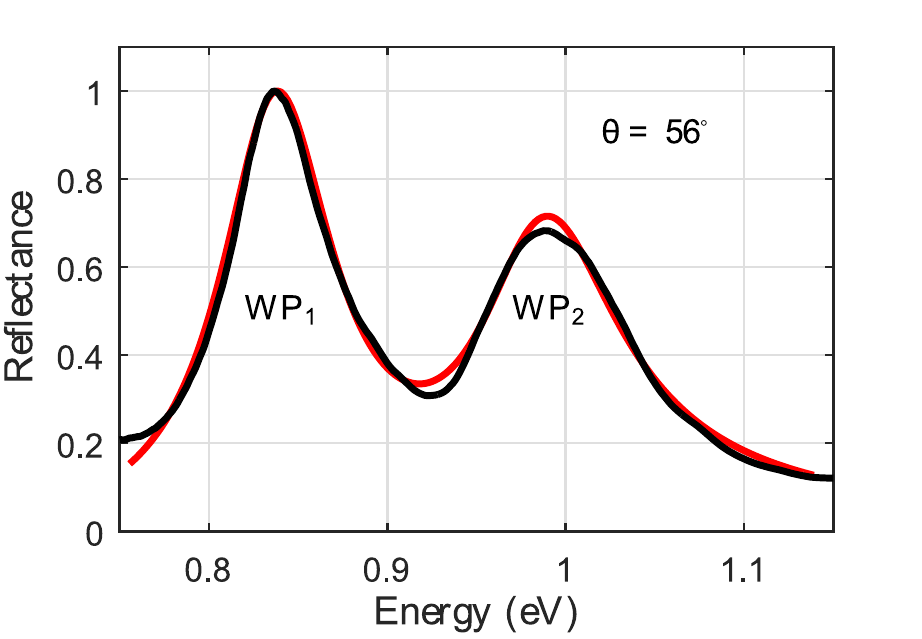}
\caption{The optical reflection spectrum (black line) for the hybrid photonic crystal(with GST film of thickness 110 nm) for the incidence angle $\theta = 56^\circ$, which demonstrates the Wood's anomalies ($WP_1$ and $WP_2$) corresponding to two close quasiguided modes. The extrapolation of the spectrum is made by equation (\ref{eq1}) (red line).}
\label{example}
\end{figure}

The analysis of the eigenmodes dispersion curves and the electric near-field distributions reveals that the peaks in the reflection spectra are attributed to the appearance of the quasiguided modes \cite{Tikhodeev2002b}. {Strictly speaking, these modes are the Fano resonances which are known to have the asymmetric line-shape \cite{Fano1941, Fano1961, luk2010fano}. In our case, the assymetry is weak and the reflection peaks can also be described in terms of a Lorentzian function. Figure\,\ref{example} shows as an example the reflection spectra for the angle $\theta = 56^\circ$ approximated by Lorentzian curves.} Since the maximum of the peaks greatly exceeds the value of the reflection coefficient at some distance from the resonances ($\max(|r_j(\omega)|^2) \gg |r_{0j}|^2$, $j=1,2$), this situation resembles the one seen in the case of the light reflection from a two-dimensional periodic array of objects demonstrating a resonant response in the absence of dielectric contrast, e.g., associated with excitons in isolated quantum dots \cite{ivchenko2000exciton, voronov2003resonance}. Relying on this analogy one can write the transmission coefficient for a hybrid structure as $t(\omega) = 1 + r(\omega)$, and taking into account the relation $|r|^2+|t|^2=1$, one gets that $|r|^2+ Re(r)=0$. Further, the reflection coefficient can be presented as a sum of the pole terms $r=\sum_{j=1}^N r_j$, $r_j=f_j/(\omega_{0j}-\omega-i\Gamma_j)$, where $\omega_{0j}$ and $\Gamma_j=\Gamma_{0j}$ are the resonant frequency and radiative damping rate of the j-th resonance. As a result, one can obtain the following expression for the reflection coefficient: $$R=|r|^2=\sum_{j=1}^N \frac{Im(f_j)\Gamma_{0j}+Re(f_j)(\omega-\omega_{0j})}{(\omega-\omega_{0j})^2 + \Gamma_{0j}^2}\:.$$By taking into consideration the absorption in the system, we can recast the last expression into the form 

\begin{equation}
R(\omega)=\sum_{j=1}^N \frac{A_j+B_j(\omega-\omega_{0j})}{(\omega-\omega_{0j})^2 + \Gamma_{j}^2}\:,
\label{eq1}
\end{equation}
where $A_j$, $B_j$ and $\Gamma_j$ are some real constants.
Alternatively, equation\,(\ref{eq1}) can be derived by considering a multipath interference from $N$ resonances.

The extrapolation of the reflection spectra shown in Fig.\,\ref{example} by equation (\ref{eq1}), where the frequency $\omega$ is expressed in terms of energy ($\omega\rightarrow E [eV]$), gives the following fitting parameters: {$A_1=0.058$, $B_1=0.077$, $\Gamma_1=0.039$ and $A_2=0.056$, $B_2=0.327$, $\Gamma_2=0.048$}. Some discrepancy in the description of the experimental spectrum can be attributed to a difference of the spectral shape for the second mode of Wood's anomaly from the Lorentzian shape (due to a large broadening of the peak) and to the frequency dependence of the phase shift due to a difference in the spatial locations of the quasiguided modes in the near-surface region.

\section{Conclusion}
In conclusion, we have suggested the model of hybrid photonic crystal structure where the vertical cross section of GST225 capping layer has the crescent-like shape. It enables us to theoretically describe the Wood's anomalies {in terms of the quasiguided modes which are localized in the gaps between silica spheres in the near-surface region.} The measured reflection spectra of such structures are in a quantitative agreement with the simulation results obtained by the scattering matrix method. One can expect that the suggested model is sufficient for prediction of optical properties other opal-based structures. 

The potential practical applications of such structures include the development of promising elements for all-optical switches, filters, and multiplexers. Although the described resonances are in the near-infrared spectral region, the sensitivity of the reflection spectra to the GST225 capping layer {paves} new way for the color management and the light control.   

\begin{acknowledgments}
The work of S.A.D., N.A.G., and S.G.T. on the theoretical description of optical response and near field distribution has been funded by Russian Science Foundation (Grant No. 16-12-10538).  A.B.P. and I.A.A. wish to thank the Russian Foundation for Basic Research (project no. 15-52-12011) and the Deutsche Forschungsgemeinschaft (DFG) within the framework of the International Collaborative Research Centre (ICRC) TRR 160. Authors kindly thank N. Feoktistov  for helping in ellipsometric measurements. The calculations were performed using the IBM high-performance cluster of the Skolkovo Institute of Science and Technology (Skoltech).
\end{acknowledgments}

%
\end{document}